\documentclass[12pt,epsf,amssymb,ulem,qsymbols]{article}
\usepackage[latin9]{inputenc}
\usepackage{amsmath}
\usepackage{amssymb}
\usepackage{graphicx}

\makeatletter

\newcommand{\bea}{\begin{eqnarray}}
\newcommand{\eea}{\end{eqnarray}}
\newcommand{\beq}{\begin{equation}}
\newcommand{\eeq}{\end{equation}}
\newcommand{\ec}{\end{center}}
\newcommand{\bc}{\begin{center}}

\newcommand{\gev}{{\rm GeV}}
\newcommand{\mev}{{\rm MeV}}
\newcommand{\kev}{{\rm keV}}
\newcommand{\pdir}{p\kern -5.2pt\raise 0.2ex\hbox {/}}

\newcommand{\vdir}{v\kern -5.75pt\raise 0.15ex\hbox {/}}
\newcommand{\kdir}{k\kern -5.75pt\raise 0.15ex\hbox {/}}
\newcommand{\epsdir}{\epsilon\kern -5.0pt\raise 0.15ex\hbox {/}}
\newcommand{\bvdir}{\bar{v}\kern -5.75pt\raise 0.15ex\hbox {/}}
\newcommand{\Ddir}{D\kern -7.75pt\raise 0.20ex\hbox {/}}
\newcommand{\Adir}{A\kern -7.75pt\raise 0.20ex\hbox {/}}
\newcommand{\ldir}{l\kern -5.0pt\raise 0.2ex\hbox{/}}
\newcommand{\varepsdir}{\varepsilon\kern -5.5pt\raise 0.15ex\hbox{/}}

\newcommand{\co}{{\cal O}}
\newcommand{\nf}{{N_{\rm f}}}

\newcommand{\nn}{\nonumber}

\makeatother

\begin{document}

\title{\textbf{\Large{}Lattice QCD estimate of the $\eta_{c}(2S)\to J/\psi\gamma$
decay rate }}

\vskip 1.2cm\par

\author{\textsc{\normalsize{}Damir Be\v{c}irevi\'{c}$^{a}$, Michael Kruse$^{b}$ and Francesco Sanfilippo$^{c}$}}
\maketitle

\par {\raggedright 
\begin{center}
{\sl
$^a$~Laboratoire de Physique Th\'eorique (B\^at.~210)~
\footnote{\textsl{Laboratoire de Physique Th\'eorique est une unit\'e mixte de recherche
du CNRS, UMR 8627.}} \\
Universit\'e Paris Sud, F-91405 Orsay-Cedex, France.}
{\par \vskip 0.25 cm\par}
{\sl 
$^b$~\'Ecole normale sup\'erieure, D\'epartement d'informatique\\
29, rue d'Ulm, F-75230 Paris, France.}
{\par \vskip 0.25 cm\par}
{\sl 
$^c$~School of Physics and Astronomy, University of Southampton\\
Southampton SO17 1BJ, UK.}
\par\end{center}
\vspace{1.4em}
\noindent\par}

\begin{abstract}
We compute the hadronic matrix element relevant to the physical radiative decay $\eta_{c}\left(2S\right)\to J/\psi\gamma$ by means of lattice QCD. 
We use the (maximally) twisted mass QCD action with $\nf =2$ light dynamical quarks and from the computations
made at four lattice spacings we were able to take the continuum limit. 
The value of the mass ratio $m_{\eta_c(2S)}/m_{\eta_c(1S)}$ we obtain is consistent with the experimental value, and our prediction for the  
form factor is $V^{\eta_{c}\left(2S\right)\to J/\psi\gamma}(0)\equiv V_{12}(0)=0.32(6)(2)$, leading to $\Gamma(\eta_c (2S) \to J/\psi\gamma)  = (15.7\pm 5.7)$~keV,  
which is much larger than  $\Gamma(\psi (2S) \to \eta_c\gamma)$ and within reach of modern experiments. 
\end{abstract}
{\small{}PACS: 12.38.Gc, 13.20.Gd, 14.40.Pq} 

\newpage{}

\section{Introduction}

Recent progress in simulations of QCD on the lattice allowed to solve
several long standing problems in hadronic physics. One such a problem
was a conflict between theoretical predictions 
of the radiative decay width $\Gamma(J/\psi\to\eta_{c}\gamma)$ and its value experimentally measured in 1986~\cite{Gaiser:1985ix}. 
That early measurement turned out to be too small
when confronted with theoretical predictions based on various quark
models~\cite{Sucher:1978wq,Grotch:1984gf,Godfrey:1985xj}, dispersion relations~\cite{Shifman:1979nx} 
and QCD sum rules~\cite{Khodjamirian:1983gd,Beilin:1984pf}.
Only in 2009 the CLEO Collaboration~\cite{Mitchell:2008aa} was able to provide a new measurement
of this decay width and found it to be over $1\sigma$ larger than the one measured
using the Crystal Ball detector in 1986~\cite{Gaiser:1985ix}, but still somewhat smaller than predicted.
A very recent measurement at the
KEDR experiment confirmed the CLEO result in that $\Gamma(J/\psi\to\eta_{c}\gamma)$ is large~\cite{Anashin:2014wva}, 
and reported a value $1.4 \sigma $ larger than that by CLEO. The charm factory at BESIII is expected to
provide a new experimental determination of this decay width and close this issue. 

On the theory side, as we just mentioned, the numerical simulations of QCD on the lattice helped solving this
problem since the corresponding form factor was computed at several
lattice spacings, thus allowing to take the continuum limit and obtain
a viable physical result. The effects of light dynamical
quarks were included in simulations. In particular, by using the maximally
twisted mass QCD on the lattice with $\nf =2$ dynamical light
quarks we confirmed in ref.~\cite{Becirevic:2012dc} the discrepancy between theory and the old measurement~\cite{Gaiser:1985ix}. 
Our finding was soon corroborated by a completely independent simulations made by the HPQCD Collaboration, 
in which the effects of $\nf =2+1$ staggered light quark flavors were included~\cite{Donald:2012ga}.
Both lattice results implemented the non-perturbative renormalization
procedure and, in the continuum limit, exhibited quite a remarkable
agreement for $\Gamma(J/\psi\to\eta_{c}\gamma)$, also in agreement with the most recent experimental findings~\cite{Anashin:2014wva}. 
Furthermore, recent improvement of
the effective theory approach, based on potential non-relativistic QCD (pNRQCD)~\cite{Brambilla:2005zw}, 
lead to a very good agreement with lattice
QCD results~\cite{Pineda:2013lta}. Therefore, as of today, the theoretical
estimate of $\Gamma(J/\psi\to\eta_{c}\gamma)$ is very solid. Another
important motivation for a more dedicated experimental study of this
decay mode lies in the fact that this decay rate could be sensitive to
the CP-odd light Higgs boson if its mass were very light, i.e.
close to that of the $\eta_{c}$-meson~\cite{A0}.

In this paper we discuss another class of so-called magnetic dipole (M1) transitions,
namely a decay of a radially excited charmonium to a ground state. In particular, we will focus on $\eta_{c}(2S)\to J/\psi\gamma$, the
width of which is yet to be measured and the lattice QCD computation of its hadronic matrix element 
should be regarded as a clear theoretical prediction. Quark model predictions of the hadronic matrix element governing this decay are
 difficult to control because the leading term
vanish due to orthogonality of the meson wave functions and the
relativistic corrections are highly sensitive to the form of the used potential. Phenomenology of this decay mode has
not been considered in the effective theory approach~\cite{Brambilla:2005zw}. It is often assumed to be similar in size to $\Gamma(\psi(2S)\to\eta_{c}\gamma)$
which has been measured and found to be small. Even if the
physical form factors $V^{\eta_c(2S)\to J/\psi\gamma}(0)$ and $V^{\psi(2S)\to\eta_c\gamma}(0)$ were equal, the width
of $\eta_{c}(2S)\to J/\psi\gamma$ would still be about three times
larger due to different spin of the initial state. However, there is a dynamical reason why $\Gamma(\psi(2S)\to\eta_{c}\gamma)$ is suppressed with respect to $\Gamma (\eta_{c}(2S)\to J/\psi\gamma)$.
It was first observed in ref.~\cite{Sucher:1978wq} that a substantial part of relativistic
corrections cancel in $\psi(2S)\to\eta_{c}\gamma$, whereas they add up in the case of $\eta_{c}(2S)\to J/\psi\gamma$ to further enhance its decay
rate. In the effective field theory approach~\cite{Brambilla:2005zw}, that point has been recently emphasized in ref.~\cite{Pineda:2013lta}, and can be compactly written as:
\begin{align}
\label{pineda}
\Gamma(\psi(2S)\to\eta_c\gamma) =
\frac{16 \alpha  }{27 m_c^2} {\widetilde q_\gamma^3}\left[
\frac{\widetilde q_\gamma^2}{24}{}_{\eta_c}\langle r^2\rangle_{\psi(2S)}+\frac{5}{6}\frac{{}_{\eta_c}\langle p^2\rangle_{\psi(2S)}}
{m_c^2} -\frac{2}{m_c^2}\frac{{}_{\eta_c}\langle V_{S^2}(\vec{r}){}\rangle_{\psi(2S)}}{E_{\psi(2S)}-E_{\eta_c}}\right]^2,\nn \\
\Gamma(\eta_c(2S)\to J/\psi\gamma) = \frac{16 \alpha  }{9 m_c^2} { q_\gamma^3} 
 \left[\frac{  q_\gamma^2}{24}{}_{J/\psi}\langle r^2 \rangle_{\eta_c(2S)}+\frac{5}{6}
\frac{{}_{J/\psi}\langle p^2 \rangle_{\eta_c(2S)}}{m_c^2} 
+\frac{2}{m_c^2}\frac{{}_{J/\psi}\langle V_{S^2}(\vec{r}) \rangle_{\eta_c(2S)}}{E_{\eta_c(2S)}-E_{J/\psi}}\right]^2,
\end{align}
where $\widetilde q_{\gamma} = |\vec {\widetilde{q}}| = (m_{\psi(2S)}^2-m_{\eta_c}^2)/(2m_{\psi(2S)})$,  
$ q_{\gamma}  = (m_{\eta_c(2S)}^2-m_{J/\psi}^2)/(2m_{\eta_c(2S)})$,  $m_c$ is the pole charm quark mass, 
and the matrix elements of $\co (1/m_c^2)$ corrections are for shortness written in the form $\langle A|\co|B\rangle\equiv {}_{A}\langle \co \rangle_{B}$.
The two decays differ in sign of the terms involving the spin dependent potential, which is a peculiarity of these, so-called hindered, M1 transitions, in contrast 
to the allowed ones (e.g. $\psi(nS)\to \eta_c (nS) \gamma$) for which the spin dependent corrections do not occur in a first few terms of the $1/m_c$-expansion. 
Assuming that ${}_{J/\psi}\langle V_{S^2}(\vec{r}) \rangle_{\eta_c(2S)} \simeq {}_{\eta_c}\langle V_{S^2}(\vec{r})\rangle_{\psi(2S)}$ and is positive,  the second decay 
width will be enhanced with respect to the first one. To compute the hadronic matrix elements of the spin dependent part of the potential one should either 
rely on quark models or attempt computing them by means of NRQCD on the lattice. In pNRQCD, in situations in which the charmonium states with principal quantum number larger than $1$ are involved, the computation of the matrix elements in eq.~(\ref{pineda}) cannot be handled analytically and should be computed by a non-perturbative method.

In this paper we will compute the hadronic matrix element relevant to $\eta_{c}(2S)\to J/\psi\gamma$
by using QCD on the lattice, without relying on NRQCD, and show that the value of the corresponding form factor is indeed significantly larger
than the one governing the $\psi(2S)\to\eta_{c}\gamma$ decay. This is the first time that such a computation is conducted and on the basis of our result
we obtain 
\begin{eqnarray}
\Gamma(\eta_{c}(2S)\to J/\psi\gamma)=(15.7 \pm 5.7)\, {\rm keV}\,.
\end{eqnarray}

The remainder of this paper is organized as follows: in sec.~\ref{sec:1} we define
the matrix element and the corresponding form factor, and discuss the strategy to ensure $q^2=0$; 
in sec.~\ref{sec:2} we discuss the two-point correlation functions and the method used 
to isolate the radially excited state the efficiency of which we test on the mass splitting between the radially excited and the lowest lying states; 
in sec.~\ref{sec:3} we describe the computation of the three-point correlation functions, extract a desired form factor and discuss the phenomenological 
consequences of our result; we conclude in sec.~\ref{sec:4}

\section{Hadronic Matrix Element\label{sec:1} }

The hadronic matrix element governing the radiative decay $\eta_c(2S)\to J/\psi\gamma^\ast$ decay can be parameterized in terms of the form factor $V_{12}(q^2)$ as,~\footnote{
For notational simplicity, in what follows we will use $V^{J/\psi\to \eta_c}(q^2) \equiv V_{11}(q^2)$, $V^{\psi(2S)\to \eta_c}(q^2) \equiv V_{21}(q^2)$, and $V^{\eta_c (2S)\to J/\psi}(q^2)\equiv V_{12}(q^2)$. }    
\begin{eqnarray}
\langle J/\psi(k,\epsilon_{\lambda})\vert J_{\mu}^{{\rm em}}\vert\eta_{c}\left(2S\right)(p)\rangle=e{\cal Q}_{c}\ \varepsilon_{\mu\nu\alpha\beta}\ \epsilon_{\lambda}^{\ast\nu}k^{\alpha}p^{\beta}\ \frac{2\ V_{12}(q^{2})}{m_{J/\psi}+m_{\eta_{c}\left(2S\right)}}\,,\label{eq:def-vector-form-factor}
\end{eqnarray}
where the relevant part of the electromagnetic current is $J_{\mu}^{{\rm em}}={\cal Q}_{c}\bar{c}\gamma_{\mu}c$,
with ${\cal Q}_{c}=2/3$ in units of $e=\sqrt{4\pi\alpha}$.
The form factor $V_{12}(q^{2})$ can be computed at various values
of $q^{2}\equiv q_{\gamma}^{2}=(p-k)^{2}$ but the one relevant to the
physical $\eta_c(2S)\to J/\psi\gamma$ decay rate (on-shell photon) should be obtained at $q^{2}=0$, viz. 
\begin{eqnarray}
\Gamma\left(\eta_{c}\left(2S\right)\to J/\psi\gamma\right) & = & \frac{64}{9}\ \frac{\alpha\  {q}_{\gamma}^{3}}{\left(m_{J/\psi}+m_{\eta_{c}\left(2S\right)}\right)^{2}}|V_{12}(0)|^{2}.\label{eq:def-decay-rate}
\end{eqnarray}
To compute the form factor $V_{12}(0)$ we proceed along the lines
discussed in ref.~\cite{Becirevic:2012dc} and compute the correlation
functions in which one of the charm quark propagators, $S_{c}(x,0)\equiv S_{c}(\vec{x},t;\vec{0},0)=\langle\bar{c}(x)c(0)\rangle$,
is computed by using the twisted boundary conditions~\cite{twbc} 
\begin{eqnarray}
S_{c}^{\vec{\theta}}(x,0;U)=e^{i\vec{\theta}\cdot\vec{x}\pi/L}S_{c}(x,0;U^{\theta})\,,\label{eq:twisted-prop}
\end{eqnarray}
where we also indicate that the propagator is computed on a gauge
field configuration 
$U_{\mu}(x)\to U_{\mu}^{\theta}(x)=e^{i\theta_{\mu}\pi/L}U_{\mu}(x)$, where
$\theta=\vartheta_{0}(0,1,1,1)$, with $\vartheta_{0}$ that should be tuned to ensure  
that $q^{2}=0$, i.e. 
\begin{eqnarray}
\vartheta_{0}=\frac{L}{\pi\sqrt{3}}\frac{m_{\eta_{c}\left(2S\right)}^{2}-m_{J/\psi}^{2}}{2m_{\eta_{c}\left(2S\right)}}\,.
\end{eqnarray}
In our previous paper we showed that the charmonium decays $J/\psi\to \ell^+\ell^-$ and $J/\psi \to \eta_c\gamma$ do not
depend on the sea quark mass. In this paper we again work with the charmonium states that are all bellow the $D^{(\ast)}\bar D^{(\ast )}$ production 
threshold and therefore the dependence on the light (sea) quark should remain negligible.~\footnote{When the energy of a charmonium state gets close to the open $D^{(\ast)}\bar D^{(\ast )}$ channel, a dependence on the light quark mass might become important~\cite{meissner}. } For that reason, in this
study, we focus on a subset of gauge field configurations considered
in ref.~\cite{Becirevic:2012dc} and study one value of the light sea quark mass 
per lattice spacing but we increase the statistics in order to be able to isolate the radially excited state from our correlation functions. 
We rely on the gauge field configurations produced by the ETM Collaboration~\cite{boucaud} in which the effect of $\nf=2$ mass-degenerate dynamical light quarks has been included by using  
the maximally twisted QCD on the lattice~\cite{fr}. Using the same action we then compute the quark propagators and correlation functions needed for the physical problem discussed in this paper. 
Details concerning the lattice ensembles and the main results of this
paper are listed in tab.~\ref{tab:parameters-and-results}.

\section{Two-point correlation functions\label{sec:2}}

\begin{table}[h!!]
\renewcommand{\arraystretch}{1.5}
\centering{}%
\begin{tabular}{|c|cccc|}
\hline 
$\beta$ & 3.80 & 3.90 & 4.05  & 4.20\\
$L^{3}\times T$ & $24^{3}\times48$  & $24^{3}\times48$  & $32^{3}\times64$ & $32^{3}\times64$ \\
$\#\ {\rm meas.}$ & 240 $\times$ 16& 552  $\times$ 16  &  750  $\times$ 16&  480  $\times$ 16 \\
\hline 
$\mu_{{\rm sea}}$ & 0.0110 & 0.0064 & 0.0030 & 0.0065\\
$a\ {\rm [fm]}$ & 0.098(3) & 0.085(3) & 0.067(2) & 0.054(1)\\
$Z_{V}(g_{0}^{2})$~\cite{ZZZ} & 0.5816(2) & 0.6103(3) & 0.6451(3) & 0.686(1)\\
$\mu_{c}$~\cite{Blossier:2010cr} & 0.2331 & 0.2150 & 0.1849 & 0.1566\\
$n_g$ &  10 & 20 & 25 & 28 \\
\hline 
$m_{\eta_{c}\left(2S\right)}/m_{\eta_{c}\left(1S\right)}$ & 1.301(5) & 1.276(8) & 1.263(18) & 1.260(16)\\
$V_{11}(q_0^2)$ & 1.330(8) &  1.447(5)&  1.544(5)& 1.616(7)\\
$V_{12}(0)$ &  0.532(21)& 0.483(29) &  0.433(51)& 0.368(60) \\
$t_{sep}$ & 20 & 20 & 26 & 26\\
\hline 
\end{tabular}\protect\caption{{\footnotesize{}\label{tab:parameters-and-results} Summary of the
 lattice ensembles used in this work (more information can be found
in ref.~\cite{boucaud}). Lattice spacings and bare charm quark masses
have been determined in ref.~\cite{Blossier:2010cr}. $\mu_{\rm sea}$ and $\mu_c$ are the bare quark masses
and are given in lattice units. $n_g$ is the smearing parameter, cf. eq.(\ref{ng}). Values of the mass ratios and the form factors 
obtained at each lattice spacing are also given. 
$q_0^2$ is specified in eq.~(\ref{q0}). 
$t_{sep}$ is the separation between 
the source operators chosen in computation of the three-point correlation functions.  $\#\ {\rm meas.}$ is written in terms of a number of independent gauge field configurations $\times$ a number of time sources used to compute propagators. }}
\end{table}

A crucial step in extraction of the matrix element between the lowest lying vector charmonium and a radially excited pseudoscalar one is to reliably project out the radially excited state. 
That is made through a careful study of two-point correlation functions which will be discussed in this section.

To compute the mass of $\eta_{c}\left(2S\right)$ we use a set of interpolating field operators, $P_{1},\dots ,P_{N}$, each  
coupling to a tower of $\bar{c}c$-states with $J^{PC}=0^{-+}$, and build a $N\times N$ matrix of two-point correlation
functions: 
\bea
C_{ij}\left(t\right)\equiv C_{P_{i},\, P_{j}}\left(t\right)\equiv\langle \sum_{\vec{x}}P_{i}^{\dagger}\left(x\right)P_{j}\left(0\right) \rangle .
\eea
Spectral decomposition of each correlation function can be written as
\bea\label{eq:0}
C_{ij}\left(t\right)=\sum_{n}\frac{{\cal Z}_{i}\left(nS\right){\cal Z}_{j}^{\ast}\left(nS\right)}{2m_{\eta_{c}(nS)}}e^{-m_{\eta_{c}(nS)}t}\,,
\eea
where the sum runs over $\eta_{c}\left(nS\right)$ states. In the above decomposition we neglected the multi\--particle states which is legitimate since we consider a 
few lowest lying states,  below the $D^{(\ast )}D^{(\ast)}$ production threshold.  ${\cal Z}_{j}\left(nS\right)$ in eq.~(\ref{eq:0}) denotes the hadronic matrix element, 
${\cal Z}_{j}\left(nS\right)\equiv\langle0\vert P_{j}\vert\eta_{c}\left(nS\right)\rangle$. 
We wish to find a linear combination of operators $P_{i}$  that couples optimally to a state $n$, viz.  $P_{\left(n\right)}=c_{i}^{\left(n\right)}P_{i}$.
The coefficients $c_{i}^{\left(n\right)}$ can be obtained by solving the Generalized Eigenvalue Problem (GEVP)~\cite{GEVP}, 
\begin{eqnarray}
 & C_{ij}(t)v_{j}^{(n)}(t,t_{0})=\lambda^{(n)}(t,t_{0})C_{ij}(t_{0})v_{j}^{(n)}(t,t_{0}), \label{eq:gevp}
\end{eqnarray}
and from the resulting eigenvectors $v_j^{(n)}$ obtain,
\bea
c_{i}^{\left(n\right)}=\left(\sqrt{C\left(t_{0}\right)}\right)_{i,j}v_{j}^{\left(n\right)}\,,\label{eq:gevp2}
\eea
our desired solution.  The parameter $t_{0}$ in eq.~(\ref{eq:gevp}) should be chosen large enough
so that the correlation functions $C_{ij}\left(t_{0}\right)$ are
dominated by the lightest $n$ states, $\eta_{c}\left(nS\right)$.
The role of $C\left(t_{0}\right)$ is to optimize the problem and help us to better isolate the lowest $n$-states. The above GEVP can be solved
for each time-slice $t,$ and the corresponding eigenvectors $v^{\left(n\right)}\left(t\right)$
are expected to be independent of $t$ when focusing onto the lowest $n$ states. 
In practice $v^{\left(n\right)}\left(t\right)$ are independent on $t$, up to the effects of statistical noise which can be reduced. 
In this paper we define $c_{i}^{\left(n\right)}$ at a particular (optimal) time-slice $t_{opt}$, 
chosen in the region where the time dependence of $c_{i}^{\left(n\right)}$
is indeed very small.

With the coefficients $c_{i}^{\left(n\right)}$ we can construct the interpolating operator $P_{\left(n\right)}=c_{i}^{\left(n\right)}P_{i}$ that couples optimally to the state $\eta_c(nS)$. 
The corresponding correlation function, 
\begin{equation}
C_{\eta_{c}\left(nS\right),\eta_{c}\left(nS\right)}\left(t\right)=\langle \sum_{\vec{x}}P_{\left(n\right)}^{\dagger}\left(x\right)P_{\left(n\right)}\left(0\right)\rangle \,,\label{eq:opt-corr}
\end{equation}
at larger values of $t$, is dominated by $\eta_{c}\left(nS\right)$, the mass of which is then extracted from
\begin{equation}
C_{\eta_{c}\left(nS\right),\eta_{c}\left(nS\right)}\left(t\right)\to \frac{{\cal Z}_{\left(n\right)}^{\ast}\left(nS\right){\cal Z}_{\left(n\right)}^{\ast}\left(nS\right)}{2m_{\eta_{c}(nS)}}e^{-m_{\eta_{c}(nS)}t}\,.\label{eq:opt-corr-td}
\end{equation}

\noindent 
In the present study we use a basis of three operators,~\footnote{Note that in the above notation we distinguished $c$ from $c^{\prime}$,
corresponding to the choice $r=+1$ and $r=-1$ in the Wilson-Dirac
operator, which is a peculiarity of twisted mass QCD action on the
lattice.
}
\begin{equation}
\begin{cases}
P_{1} & =\bar{c}\gamma_{0}\gamma_{5}\gamma_{i}\nabla_{i}c^{\prime}\\
P_{2} & =\bar{\mathbf{c}}\gamma_{5}\mathbf{c^{\prime}}\\
P_{3} & =\bar{\mathbf{c}}\gamma_{0}\gamma_{5}\gamma_{i}\nabla_{i}\mathbf{c^{\prime}}
\end{cases}\,,\label{eq:used_set}
\end{equation}
where the symmetric
covariant derivative on the lattice is defined as:
\begin{align} \label{eq:covder}
\nabla_{i}f\left(n\right)=\frac{1}{2}\left(U_{n;i}^{n_{a}}f_{n+\hat{i}}-U_{n-\hat{i};i}^{n_{a}\dagger}f_{n-\hat{i}}\right)\,.
\end{align}
Note that in eq.~(\ref{eq:used_set}) we use $\mathbf{c}=\mathcal{H}c$, to distinguish the smeared quark field from the local one, 
with the smearing operator $\mathcal{H}$ given by~\cite{Gusken:1989ad}:
\begin{align}\label{ng}
\mathcal{H}=\left(\frac{1+\kappa H}{1+6\kappa}\right)^{n_{g}}\,,
\end{align}
\begin{align}
H_{n,m}=\sum_{i=1}^{3}\left(U_{n;i}^{n_{a}}\delta_{n+\hat{i},m}+U_{n-\hat{i};i}^{^{n_{a}}\dagger}\delta_{n-\hat{i},m}\right)\,.
\end{align}
The links $U_{n;i}^{n_{a}}$ entering the smearing operator and the covariant derivative~(\ref{eq:covder}) are $n_{a}$-times 
APE smeared~\cite{albanese}, i.e. they are obtained from the $(n_{a}-1)$-times smeared link $U_{n;i}^{(n_{a}-1)}$ and its surrounding {\em staples},
denoted by $V_{i,\mu}^{(n_{a}-1)}$, namely 
\begin{equation}
U_{n;i}^{n_{a}}={\rm Proj_{SU(3)}}\left[\left(1-\alpha\right)U_{n;i}^{\left(n_{a}-1\right)}+\frac{\alpha}{6}V_{n;i}^{\left(n_{a}-1\right)}\right]\,.
\end{equation}
In the present study we use $\alpha=0.5$ and $n_a=20$ for all our lattices. 
The other smearing parameters that appear in $\mathcal{H}$ are $n_g$ and $\kappa$. We keep $\kappa = 4$ for all our lattices, 
but the number of steps $n_g$ as given in tab.~\ref{tab:parameters-and-results}.

With the above set of operators~(\ref{eq:used_set}) we focus on the first three states, our main target being the radial excitation $\eta\left(2S\right)$. We checked that the 
inclusion of more operators, in a way it has been done in e.g. ref.~\cite{Liu:2012ze},  does not lead to any improvement of the signal for the state $\eta\left(2S\right)$. With the statistical quality of our data more operators included in GEVP would not help improving the extraction of higher excited states either. 
The values of coefficients $c_{1,2,3}^{(n)}$ we find for the first two states, as well as the values we take for $t_0$ and $t_{opt}$, are given in tab.~\ref{tab:coeffs}.
\begin{table}
\renewcommand{\arraystretch}{1.5}
\centering{}%
\begin{tabular}{|c|cccc|}
\hline 
$\quad \beta\quad $ & 3.80 & 3.90 & 4.05  & 4.20\\  \hline
$c_1^{(1)}$ & $0.1339(8)$  & $0.0755(4)$  & $0.0619(3)$ & $0.2259(12)$ \\ 
$c_2^{(1)}$ & $0.8247(13)$  & $0.8900(9)$  & $0.9313(6)$ & $0.7739(13)$ \\
$c_3^{(1)}$ & $0.0414(6)$  & $0.0344(5)$  & $0.0068(2)$ & $0.0003(1)$ \\
\hline 
$c_1^{(2)}$ & $0.390(9)$  & $0.217(6)$  & $0.150(2)$ & $0.313(7)$ \\
$c_2^{(2)}$ & $0.161(2)$  & $0.090(2)$  & $0.0313(7)$ & $0.1001(34)$ \\
$c_3^{(2)}$ & $0.449(12)$  & $0.694(8)$  & $0.819(3)$ & $0.587(10)$ \\
\hline 
$t_0$ & $1$  & $1$  & $2$ & $6$ \\ 
$t_{opt}$ & $6$  & $8$  & $9$ & $11$ \\   \hline
\end{tabular}\protect\caption{{\footnotesize{}\label{tab:coeffs} Coefficients $c_i^{(n)}$ determined 
by solving the GEVP in eq.~(\ref{eq:gevp}) in the basis of operators listed in eq.~(\ref{eq:used_set}). $n=1$ refers to the optimal coupling to the lowest lying state, and 
$n=2$ to its first radial excitation. We also give the values of $t_0$ and $t_{opt}$ that we chose while solving the GEVP [cf. text following eqs.~(\ref{eq:gevp},\ref{eq:gevp2})].}}
\end{table}

\begin{center}
\begin{figure}[t!]
\begin{centering}
\includegraphics[width=0.5\textwidth]{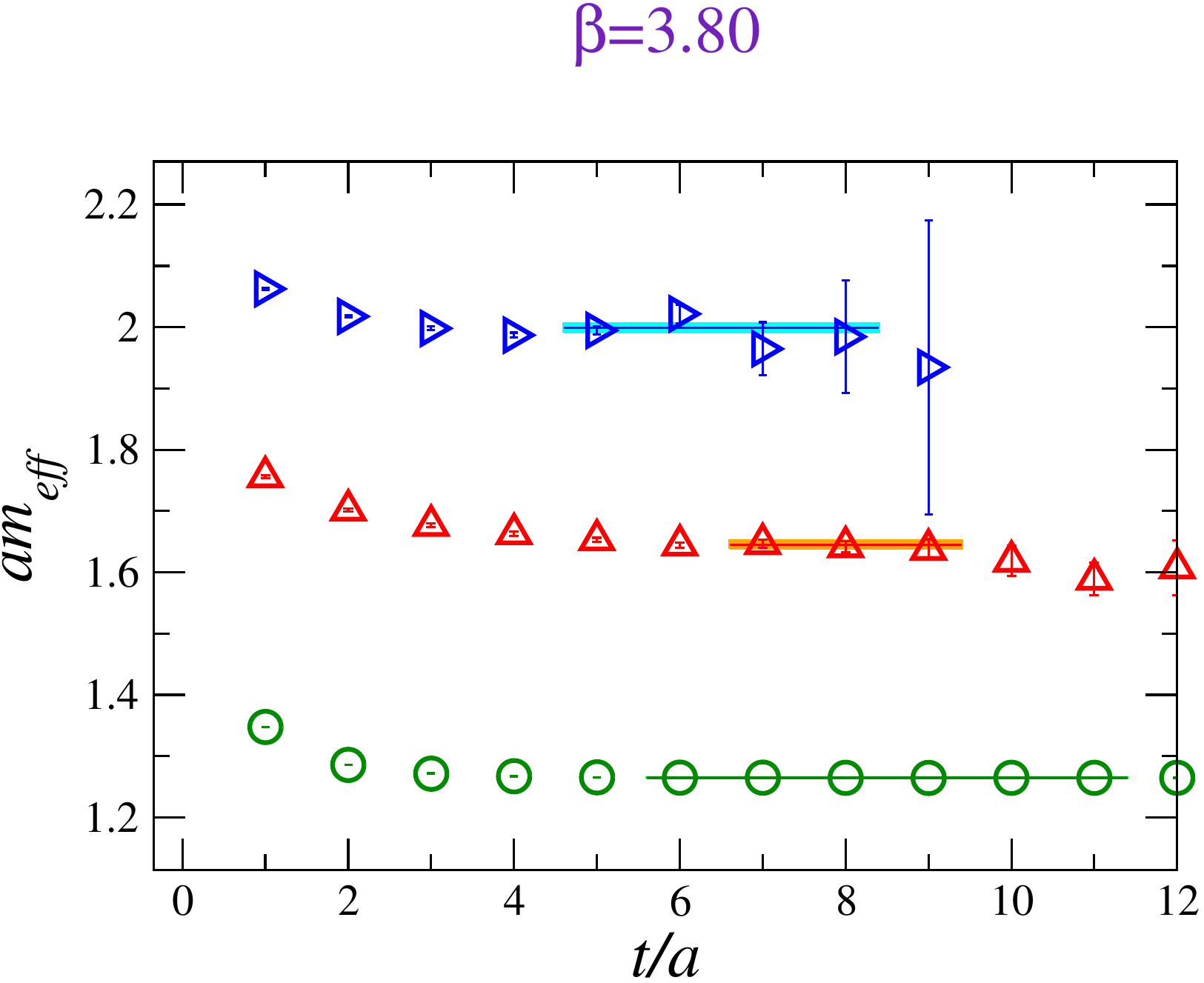}\includegraphics[width=0.5\textwidth]{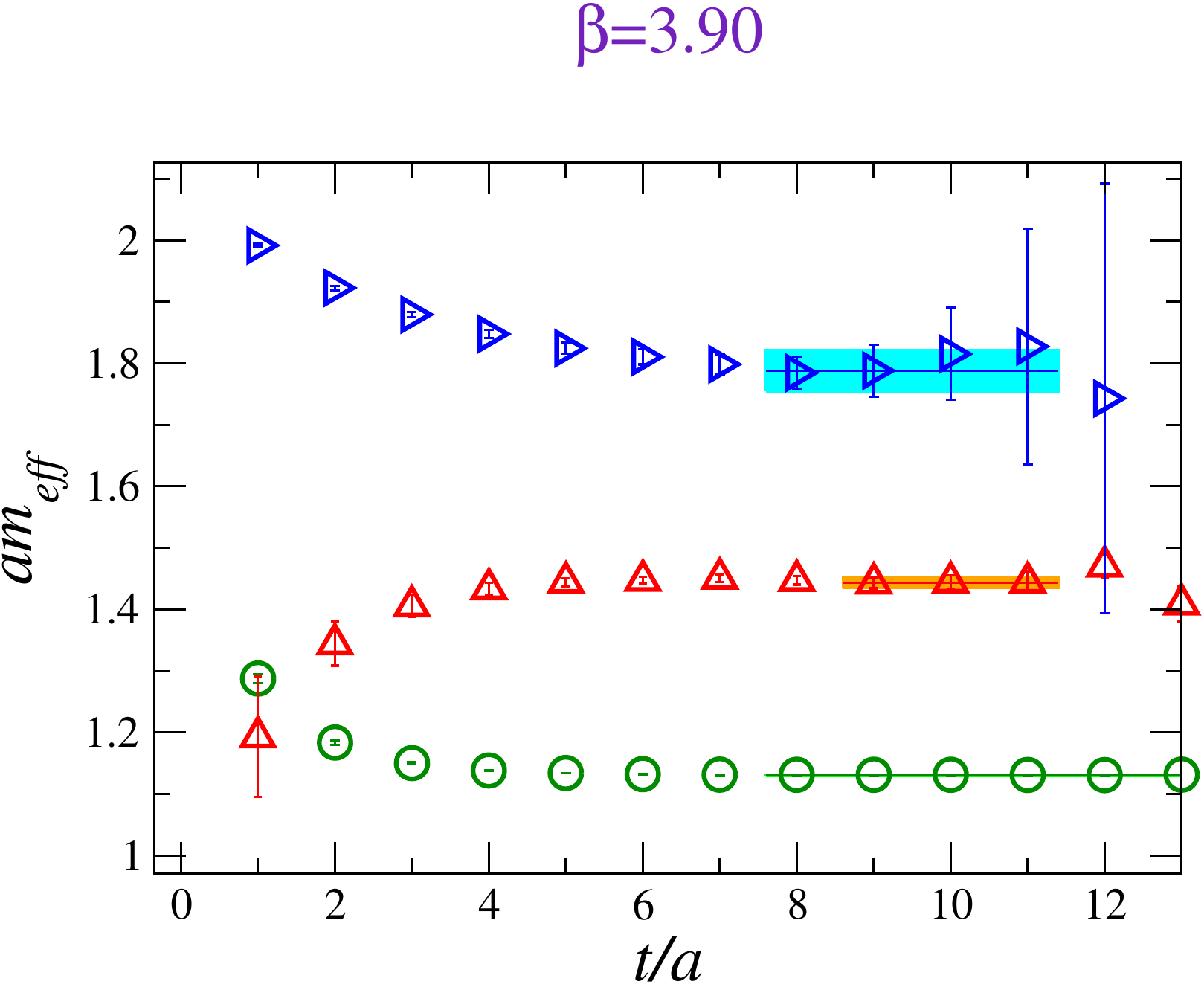}
\par\end{centering}

\begin{centering}
\includegraphics[width=0.5\textwidth]{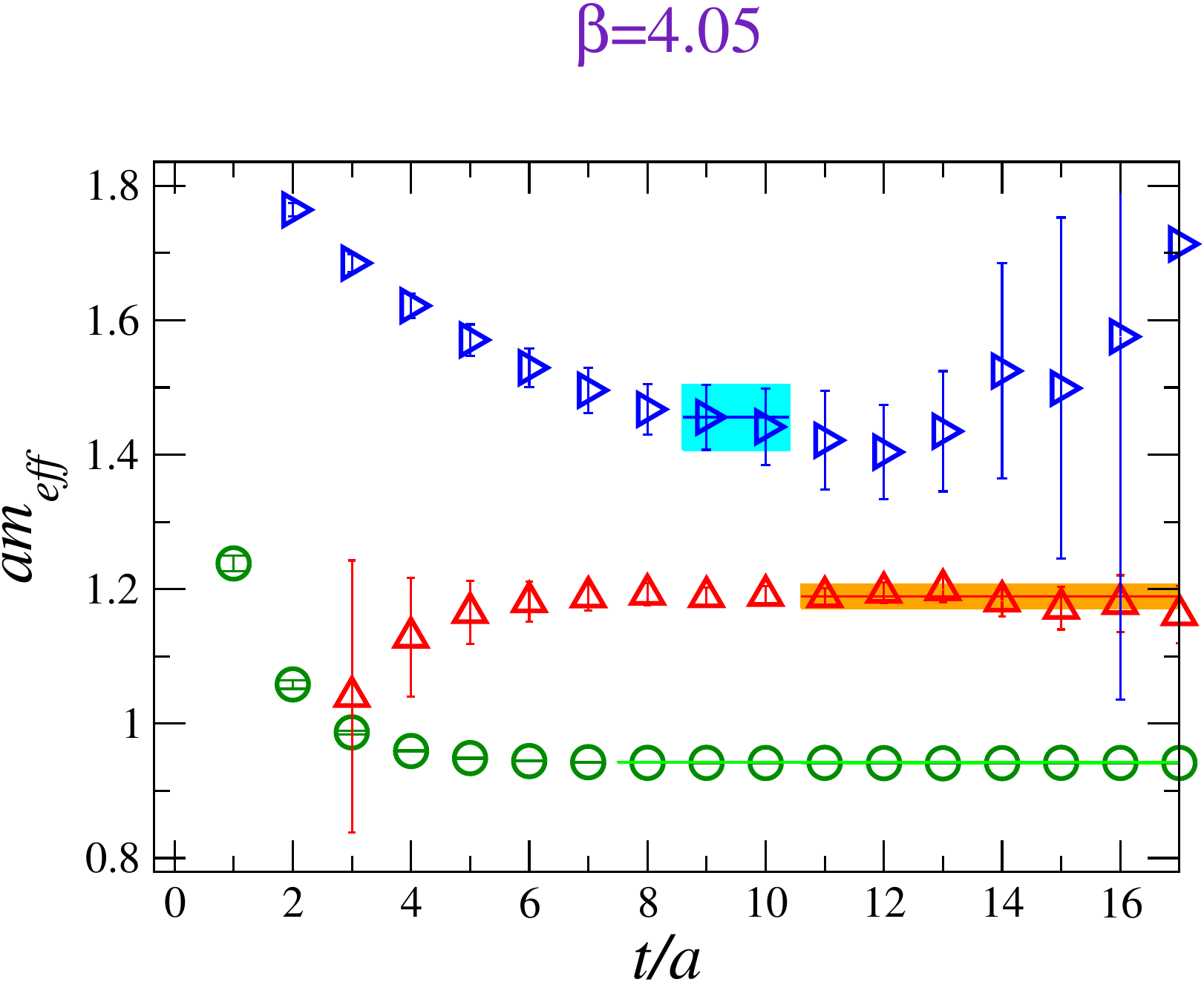}\includegraphics[width=0.5\textwidth]{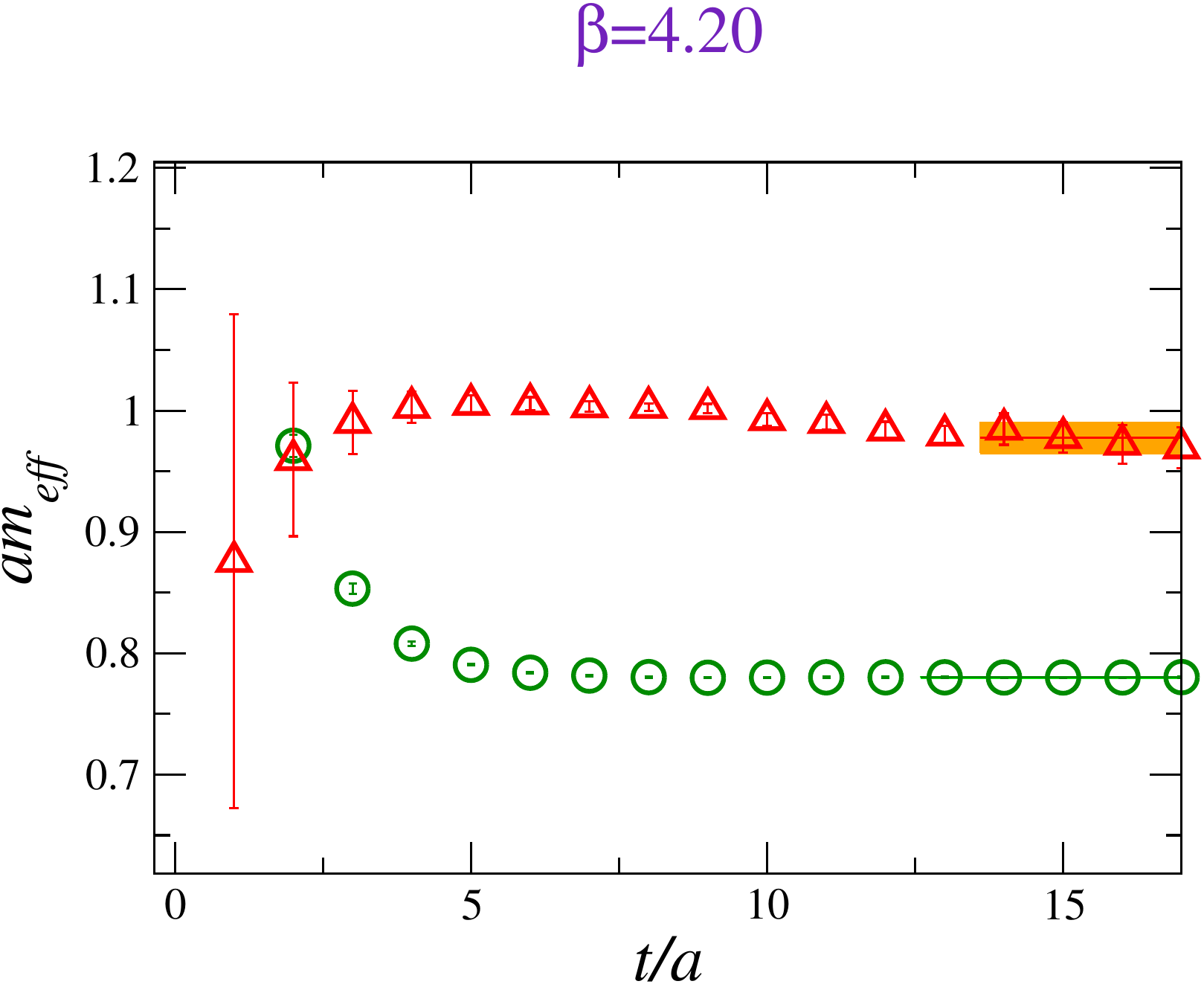}
\par\end{centering}

\centering{}\protect\caption{\label{fig:eff-mass-gevp}\textsl{\footnotesize{}Effective masses
of the charmonium states, $m_{\eta_{c}\left(1,2,3S\right)}^{{\rm eff}}(t)$, extracted
from the two-point correlation functions according to eq.~(\ref{eq:meff})
at four lattice spacings. The bands display the masses resulting from a 
fit to constant over the corresponding fit range.}}
\end{figure}

\par\end{center}

To determine simultaneously the mass $m_{\eta_{c}\left( n S\right)}$
and the matrix element $\mathcal{Z}_{\left(n\right)}\left(nS\right)$
we fit the correlation functions $C_{\eta_{c}\left(nS\right),\eta_{c}\left(nS\right)}\left(t\right)$,
defined in eq.~(\ref{eq:opt-corr}), over an appropriate time interval
using eq.~(\ref{eq:opt-corr-td}). In fig.~\ref{fig:eff-mass-gevp}
we show the effective mass $m_{\eta_{c}\left(nS\right)}^{{\rm eff}}\left(t\right)$
defined as:~\footnote{This definition is valid as long as we consider correlation function
at time-slices far enough from $T/2$, so that we can safely ignore the 
back-propagating signal.}
\begin{eqnarray}
m_{\eta_{c}\left(nS\right)}^{{\rm eff}}\left(t\right) & = & \log\frac{C_{\eta_{c}\left(nS\right),\eta_{c}\left(nS\right)}\left(t\right)}{C_{\eta_{c}\left(nS\right),\eta_{c}\left(nS\right)}\left(t+1\right)}\,,\label{eq:meff}
\end{eqnarray}
together with the results of the fits to a constant for three lowest lying states, and for all four lattice spacings discussed in this work. Note that at $\beta=4.2$ the statistical quality of our data did not allow us to distinguish the second radial excitation. The mass of the lowest lying state is improved with respect to the results presented in our previous papers~\cite{Becirevic:2012dc,Becirevic:2013bsa}, but the overall error bars remain the same since it is entirely dominated by the error in lattice spacing. Instead of looking for absolute values of the meson masses, we prefer to compute the ratio of the radial excitation with respect to the ground state, thus eliminating the error on lattice spacing from the discussion. 
In fig.~\ref{fig:eff-mass-gevp} we show the plateaux for
the first two states that are pronounced  and of good quality. After fitting each effective mass to a constant we were able to extract $m_{\eta_{c}\left(2S\right)}/m_{\eta_{c}\left(1S\right)}$, in an obvious notation $\eta_c\equiv \eta_c$. The results are reported in tab.~\ref{tab:parameters-and-results}. 

\begin{figure}[t!]
\centering{}\includegraphics[width=0.75\textwidth]{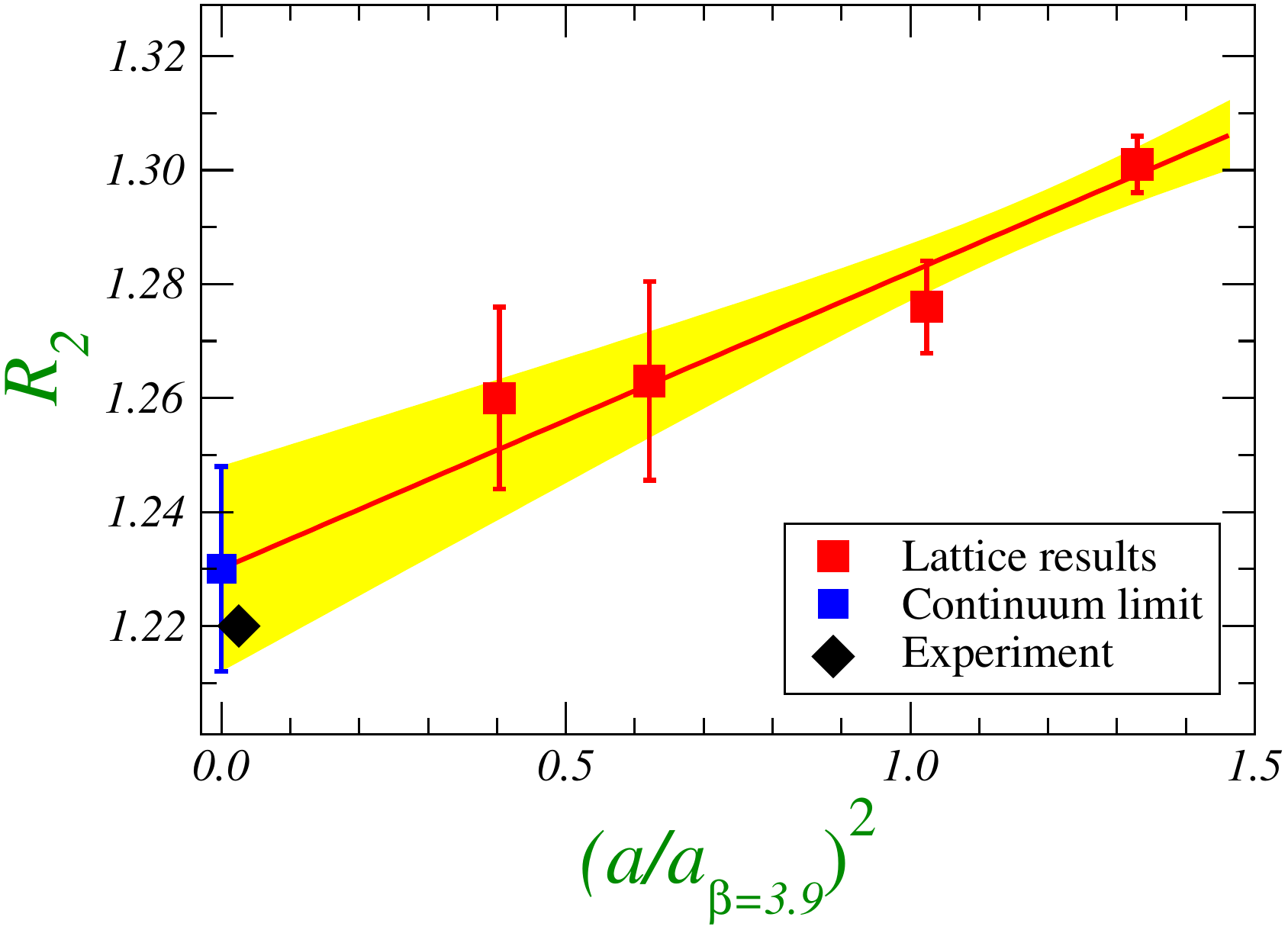}\protect\caption{\label{fig:cont-extr-r2}{\footnotesize{} Extrapolation of $R_2(a)$ to the continuum limit by using eq.~(\ref{eq:ratios-meta2c-over-meta1c}). 
Points from right to left corresponds to $\beta=\{3.80,\,3.90,\,4.05,\,4.20\}$,
the solid line and band show the fit result with its error.}}
\end{figure}

Strictly speaking $m_{\eta_c}$ is not a lattice result. It is just a cross-check because the mass of the charm quark ($\mu_c$ in tab.~\ref{tab:parameters-and-results}) has been tuned in ref.~\cite{Blossier:2010cr} in such a way as to reproduce the correct $m_{\eta_{c}}^{exp}=2980.3\, \mev$, and was then checked to result in a correct physical $m_{D_{(s)}}$ in the continuum limit. The results for $m_{\eta_{c}\left(2S\right)}/m_{\eta_{c}}$, instead, are clean lattice QCD results. To get a physically relevant result we need to make the continuum extrapolation, which we do by using 
\begin{equation}
R_2 (a)\equiv \frac{m_{\eta_{c}\left(2S\right)}\left(a\right)}{m_{\eta_{c}}\left(a\right)} = R_2^{\rm cont}\left[ 1 + X_R \left({a\over a_{(\beta=3.9)}}\right)^2
\right]\,,\label{eq:ratios-meta2c-over-meta1c}
\end{equation}
where we account for the dominant ${\cal O}(a^2)$ discretization effects~\cite{fr}. The above form appears to be adequate to describe our data, and the result of that extrapolation 
is shown in fig.~\ref{fig:cont-extr-r2}. 
We obtain
\bea
R_2^{\rm cont} = 1.230(18), \qquad X_R = 0.042(13)\,,
\eea 
in very good agreement with the experimentally established $R_2^{\rm exp}=1.220(1)$~\cite{PDG}. 
The parameter $X_R$ measures the shift of the continuum value with respect to the one obtained at the lattice with $a=0.085(3)$~fm, which appears to be in the range of 
$3\div 5$\%. The above quoted errors are statistical only. By modifying (enlarging) the plateau region and including $2$ more points, we end up with the fully compatible results, which then in the continuum limit give $R_2^{\rm cont} = 1.226(18)$. In view of the fact that our lattice QCD result has a much larger error than the corresponding physical result, we will not further dwell on systematics but simply conclude that the lattice results obtained by solving the GEVP are adequately described by eq.~(\ref{eq:ratios-meta2c-over-meta1c}) and the result obtained in the continuum limit is fully compatible with the physical  $R_2^{\rm exp}=1.220(1)$. 

This is not the first lattice determination of $R_2$ but it is the first in which the maximally twisted mass QCD on the lattice has been used for its computation. In~ref.~ \cite{DeTar:2012xk} the authors obtained a slightly larger value for $R_2$ and argued that a possible source of discrepancy could be attributed to the vicinity of the $D^{(*)}D$-thresholds which they studied in the case of spin averaged $c\bar c$-states. 
That cannot be a problem in our study since we focus on the pseudoscalar state and the first open channel would be $D D^\ast$ with $D^\ast$ in its $P$-wave, which for a periodic lattice box of size $L\simeq 2$~fm results in an energy well above masses of the first few $\eta_c (nS)$ states.  
The Hadron
Spectrum Collaboration focused on the mass difference $m_{\eta_{c}\left(2S\right)}-m_{\eta_{c}\text{\ensuremath{\left(1S\right)} }}$ and obtained $663(3)\,{\rm MeV}$, in agreement with the experimental
result $658(1)\,{\rm MeV}$, despite the fact that they worked at one lattice spacing only~\cite{Liu:2012ze}.

Another important observation comes from the comparison of our result, $R_2^{\rm cont} = 1.230(18)$, with the physical  $R_2^{\rm exp}=1.220(1)$. In that respect the inclusion of non-local operators 
$P_{1}$ and $P_{3}$ in the set of operators used to solve the GEVP in eq.~(\ref{eq:used_set}) is crucial. 
In a preliminary study we used only a set of $P_2$ operators that differ between each other by a choice of smearing parameters. Despite the fact that we optimized the smearing parameters in a way that the coupling to $\eta_c(2S)$ is larger/smaller, the resulting splitting between
$\eta_{c}\left(2S\right)$ and $\eta_{c}\left(1S\right)$, in the continuum limit, was much larger (by about $250\ \mev$) than the physical one. 
That observation depends on the physical quantity we consider. For example, the results for the form factor $V_{12}(q^2)$ are more robust and remain fully compatible in both situations: (i) in which we use the operator basis~(\ref{eq:used_set}), (ii) when the basis consists of $P_2$ operators only, differing from each other by the amount of smearing implemented.   

\section{Transition Form Factor for $\eta_{c}\left(2S\right)\rightarrow J/\psi\gamma$ \label{sec:3}}

We now turn to the extraction of the form factor relevant to $\eta_{c}\left(2S\right)\rightarrow J/\psi\gamma$. We first compute the three-point correlation function, 
\begin{equation}
C_{ij}(\vec{q};t)=\sum_{\vec{x},\vec{y},\vec{z}}\langle P_{\left(2\right)}(\vec{x},0)J_{j}^{{\rm em}}(\vec{y},t)V_{i}^{\dagger}(\vec{z},t_{sep})\rangle\ e^{i\vec{q}\cdot(\vec{y}-\vec{x})} , \label{eq:def-threee-pts}
\end{equation}
where $P_{\left(2\right)}$ is the operator optimally interpolating
the $\eta_{c}\left(2S\right)$ state obtained in the previous section, $V_{i}=\bar{\mathbf{c}}\gamma_{i}\mathbf{c}^{\prime}$
is the smeared operator interpolating the $J/\psi$, and $J_{j}^{em}=\bar{c}\gamma_{j}c$
is the local vector current, renormalized by using $Z_{V}(g_{0}^{2})$
given in tab.~\ref{tab:parameters-and-results}. We benefit from the time reversal symmetry that relates the photon emission and the photon absorption processes which in terms of our correlation functions means, 
\begin{align}\label{eq:Tsym}
\sum_{\vec{x},\vec{y},\vec{z}}\langle V_{i}(\vec{z},t_{sep})J_{j}^{{\rm em}}(\vec{y},t)P_{\left(2\right)}^{\dagger}(\vec{x},0)\rangle\ e^{i\vec{q}\cdot(\vec{y}-\vec{x})}=\sum_{\vec{x},\vec{y},\vec{z}}\langle P_{\left(2\right)}(\vec{z},t_{sep})J_{j}^{{\rm em}}(\vec{y},t)V_{i}^{\dagger}(\vec{x},0)\rangle\ e^{i\vec{q}\cdot(\vec{y}-\vec{x})}\,.
\end{align}
Since the computation of  correlation function for the latter process requires less propagator inversions, in the following we will discuss the right hand side of eq.~(\ref{eq:Tsym}). The corresponding Wick contraction  reads,
\begin{align}\label{eq:wick}
{\rm Tr}\biggl[ \mathcal{P}_{t_{sep}}\left(\vec{z},\vec{z}^{\prime}\right)S_{c}\left(\vec{z}^{\prime},t_{sep};\,\vec{y},t\right)\gamma_{j}S_{c}\left(\vec{y},t;\,\vec{x},0\right)\mathcal{V}_{t}\left(\vec{x},\vec{x}^{\prime}\right)S_{c}^{\theta}\left(\vec{x}^{\prime},0;\,\vec{z},t_{sep}\right)\biggr]\,,
\end{align}
where, instead of explicitly injecting the momentum $\vec q$ to the correlation function, we use the twisted boundary condition on one of the charm quark propagators labelled by the superscript `$\theta$'. In our computation of the quark propagator $S_{c}(x;y)$ we use the stochastic source technique described in ref.~\cite{boucaud}. 
In the present study we neglect the disconnected contractions arising in the computation of all correlation functions, which is equivalent to studying these processes in a theory that contains a doublet of charm quarks  and we focussed on the non-singlet states. Such an approximation is expected to have a small impact on physical observables which is what we observed in our previous paper~\cite{Becirevic:2012dc} (see also ref.~\cite{Donald:2012ga}).

Following the discussion made in sec.~\ref{sec:2}, the explicit expressions of the interpolating field operators relevant to $\eta_c(2S)$ and $J/\psi$ states, at a given time-slice $t$, are:
\begin{align}\label{eq:P}
\mathcal{P}\left(\vec{x},\vec{x}^{\prime}\right)= & c_1^{(2)}\gamma_{0}\gamma_{5}\gamma_{i}\nabla_{i}\left(\vec{x},\vec{x}^{\prime}\right)\nn\\
&+\mathcal{H}\left(\vec{x},\vec{x}^{\prime\prime}\right)\left[c_2^{(2)}\gamma_{5}\delta\left(\vec{x}^{\prime\prime},\vec{x}^{\prime\prime\prime}\right)+c_3^{(2)}\gamma_{0}\gamma_{5}\gamma_{i}\nabla_{i}\left(\vec{x}^{\prime\prime},\vec{x}^{\prime\prime\prime}\right)\right]\mathcal{H}\left(\vec{x}^{\prime\prime\prime},\vec{x}^{\prime}\right), \nn\\
&  \nn\\
\mathcal{V}^{i}\left(\vec{x},\vec{x}^{\prime}\right)= & \mathcal{H}\left(\vec{x},\vec{x}^{\prime\prime}\right)\gamma_{i}\mathcal{H}\left(\vec{x}^{\prime\prime},\vec{x}^{\prime}\right)\ ,
\end{align}
where the coefficients $c_2^{(2)}$ are the same ones we discussed in sec.~\ref{sec:2}, and listed in tab.~\ref{tab:coeffs}. For $0\ll t\ll T/2$ the correlation function~(\ref{eq:def-threee-pts}) is dominated by the signal corresponding to $\langle \eta_{c}\left(2S\right) \vert J_{\mu}^{{\rm em}}\vert J/\psi\rangle$, i.e.
\begin{align}
C_{ij}\left(\vec{q},t\right)\simeq \frac{{\cal Z}_{P_{\left(2\right)}}{\cal Z}_{V}}{4E_{J/\psi}m_{\eta_{c}\left(2S\right)}}\exp\left[-E_{\eta_{c}\left(2S\right)}t-m_{J/\psi}(T/2-t)\right]\langle\eta_{c}\left(2S\right)\vert J_{j}^{{\rm em}}\vert J/\psi(\vec{q},\epsilon_{i})\rangle\,,
\end{align}
 where $\eta_c(2S)$ is at rest, and the couplings $\mathcal{Z}$ are given by: 
\begin{align}
{\cal Z}_{P_{\left(2\right)}} & =  \langle\eta_{c}\left(2S\right)\vert P_{\left(2\right)}\vert0\rangle \,,\nn \\
\varepsilon_{i} {\cal Z}_{V}(\vec{q}) & =  \frac{1}{3}\sum_{i=1}^{3} \langle J/\psi\left(\vec{q},\,\varepsilon_{i}\right)\vert V_{i}\vert0\rangle \,.
\end{align}
In practice, $t_{sep}=T/2$ that was suitable for the study of $J/\psi \to \eta_c\gamma$, may be too large for extraction of the matrix element involving a radially excited state. To increase the region in which we can extract the 
matrix element  $\langle\eta_{c}\left(2S\right)\vert J_{j}^{{\rm em}}\vert J/\psi(\vec{q},\epsilon_{i})\rangle$
we choose $t_{sep}<T/2$. The values of $t_{sep}$ are also given in tab.~\ref{tab:parameters-and-results}. 
Furthermore, since we take our three-momentum to be isotropic, $\vec{q}=(1,1,1)\times\vartheta_{0}\pi/L$,
we can average over six non-zero contributions, namely,
\begin{eqnarray}
C_{V}(\vec{q};t) & = & \frac{1}{6}\sum_{i=1}^{3}\epsilon_{ijk}C_{jk}(\vec{q};t)\,.\label{eq:averaged-three-pts}
\end{eqnarray}

\begin{figure}
\begin{centering}
\includegraphics[width=0.75\textwidth]{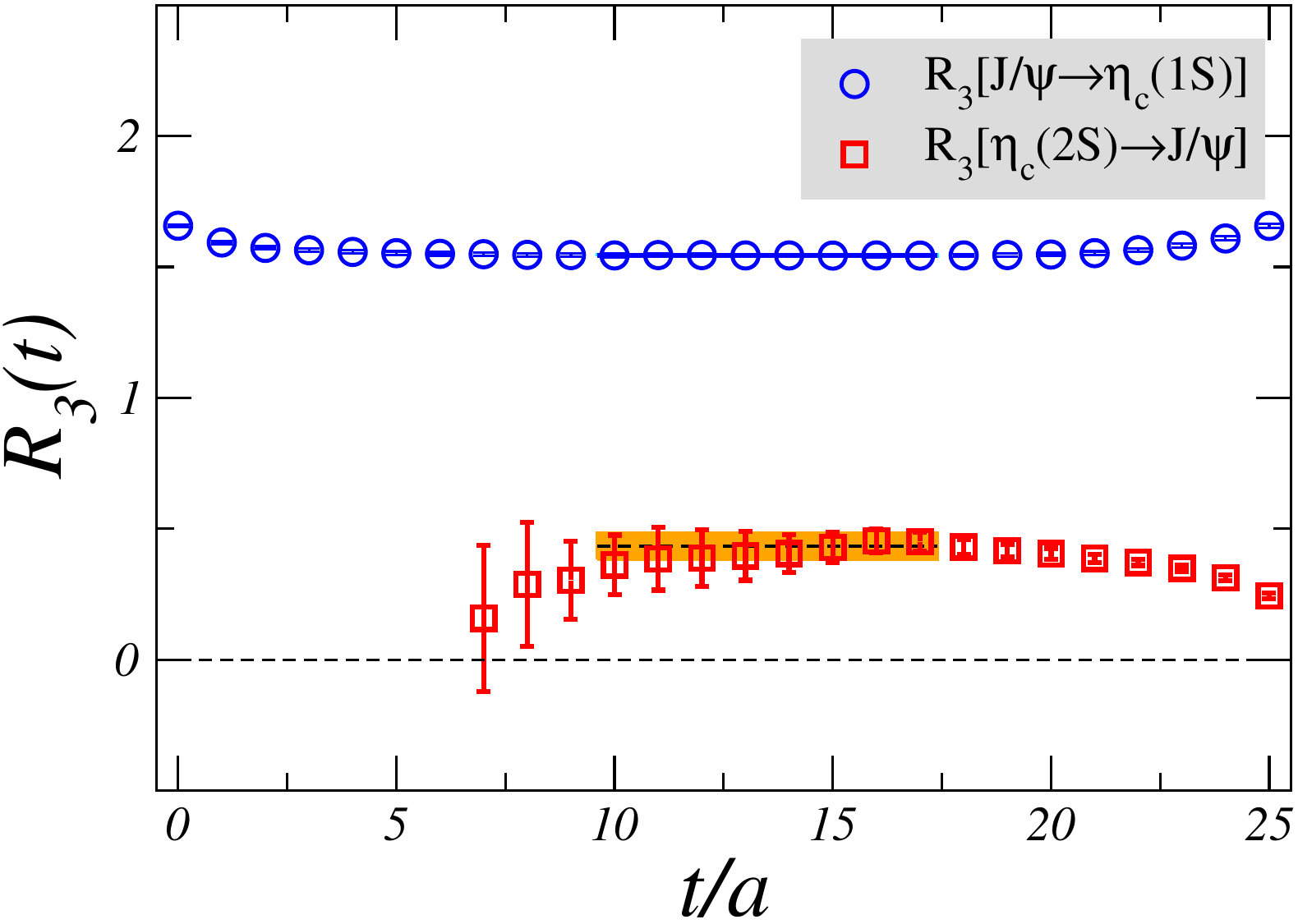}
\par\end{centering}
\centering{}\protect\caption{\label{fig:ratio_plateaux}{\footnotesize{}Plateaus of the
ratio $R_3(t)$ defined in eq.~(\ref{eq:ratio}) and obtained on the lattice with $\beta=4.05$.
Illustrated are both transitions, $\eta_c(2S) \to J/\psi \gamma$ and  $\eta_{c}\left(1S\right)\to J/\psi \gamma^\ast$.}}
\end{figure}

The matrix element is then obtained after dividing the source operators from the three-point function~(\ref{eq:def-threee-pts}), namely, 
\begin{eqnarray}
R_3(t)=\frac{C_{V}(\vec{q};t)}{{\cal Z}_{P_{\left(2\right)}}{\cal Z}_{V}\left(\vec{q}\right)}\times4E_{J/\psi}m_{\eta_{c}\left(2S\right)}\ e^{E_{J/\psi}t+m_{\eta_{c}\left(2S\right)}(t_{sep}-t)}\,,\label{eq:ratio}
\end{eqnarray}
where the values of ${\cal Z}_{V,P_{\left(2\right)}}$, $E_{J/\psi}$ and $m_{\eta_{c}}$ are obtained from the study of two-point correlation functions. 
In fig.~\ref{fig:ratio_plateaux} we illustrate the plateau of $R_3(t)$ which is then fit to a constant (shaded area in fig.~\ref{fig:ratio_plateaux}) that corresponds to the matrix element, $\langle J/\psi \vert J_{\mu}^{{\rm em}}\vert\eta_{c}\left(2S\right)\rangle$, from which we then get the form factor $V_{12}(0)$, c.f. eq.~(\ref{eq:def-vector-form-factor}). 
Furthermore, by replacing the coefficients $c_{1,2,3}^{(2)}\to c_{1,2,3}^{(1)}$ in eq.~(\ref{eq:P}) and  from the same correlation function~(\ref{eq:wick}), we also get the matrix element relevant to 
$J/\psi \to \eta_c\gamma^\ast$ decay, i.e. the form factor $V_{11}(q_0^2)$. Since the value of $\vartheta_0$ has been chosen to ensure that the emitted photon in $\eta_c(2S)\to J/\psi \gamma$ 
is on shell ($q^2=0$), after a trivial algebra one gets 
\bea\label{q0}
q_0^2 = m_{\eta_c}^2   \left[   1 - R_2 +  R_{J/\psi}^2 \left( 1 - \frac{1}{R_2}\right)  \right]\,, 
\eea
which is convenient because $m_{\eta_c}$ is in our study used as input to fix the value of the bare charm quark mass, and therefore $m_{\eta_c}$ computed on each of our lattices is equal to the physical $m_{\eta_c}^{\rm exp}=2.981(1)$~GeV by construction. Instead, $R_{J/\psi}=m_{J/\psi}/m_{\eta_c}$, discussed in our previous paper~\cite{Becirevic:2012dc}, and  $R_{2}=m_{\eta_c(2S)}/m_{\eta_c}$, computed in this work, vary with lattice spacing, and therefore $q_0^2$ also changes from one lattice spacing to another. We checked that by using $V_{11}(0) = V_{11}(q_0^2) \exp[|q_0^2|/(16 b^2)]$, with $b=0.54(1)$~GeV or $b=0.58(2)$~GeV, as found in refs.~\cite{lattice-radiative-1} and \cite{lattice-radiative-2} respectively, we reproduce our results for $V_{11}(0)$ presented in ref.~\cite{Becirevic:2012dc}.

\begin{figure}
\begin{centering}
\includegraphics[width=0.73\textwidth]{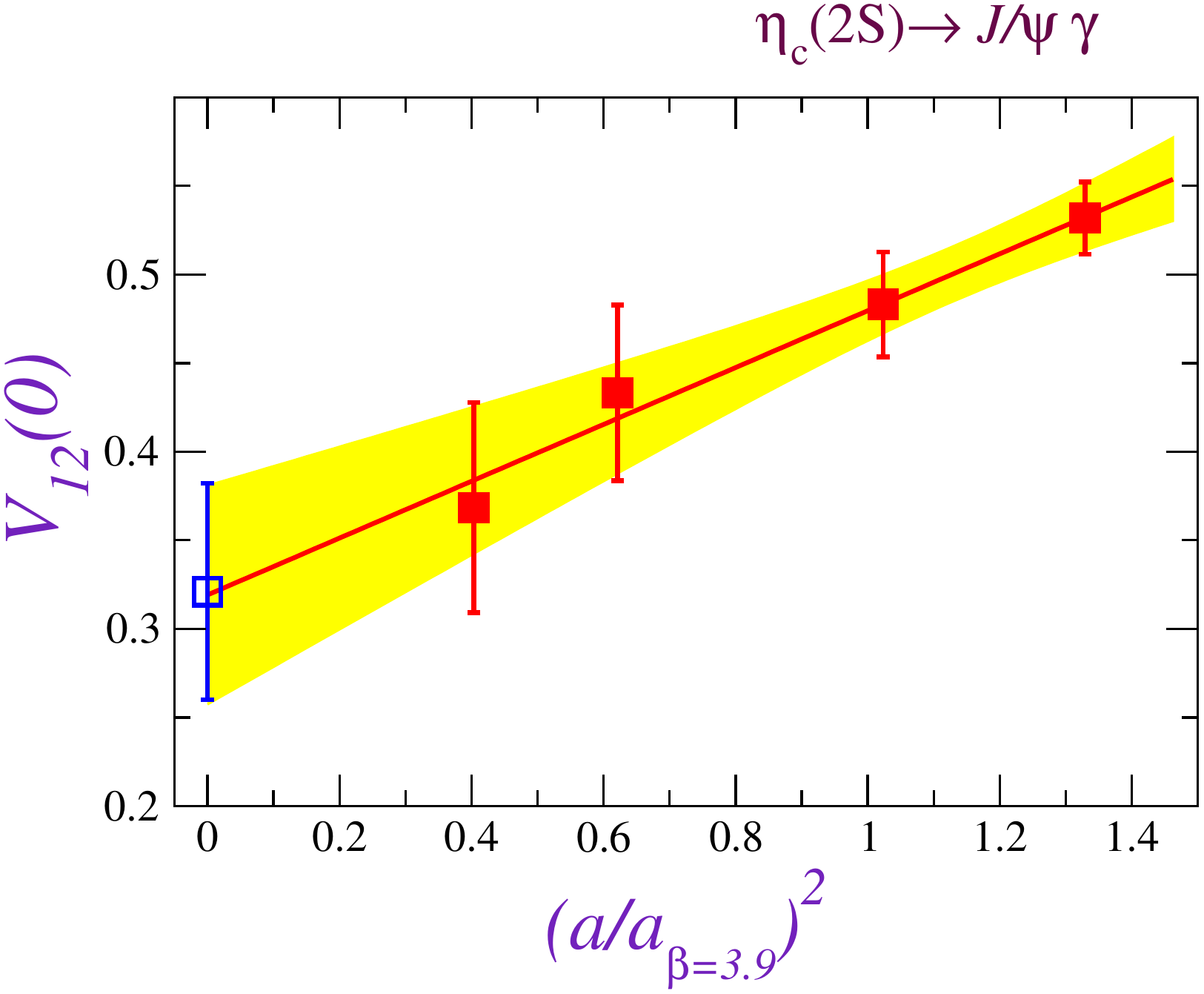}
\par\end{centering}
\centering{}\protect\caption{\label{fig:extrapo-V}{\footnotesize{}Continuum extrapolation of
the form factor $V_{\left(2\right)}(0)$ computed on our lattices
at 4 lattice spacings. The continuum extrapolation is made according
to eq.~(\ref{extrap-V}).}}
\end{figure}

Finally, we need to extrapolate our results for $V_{12}(0)$ to the continuum limit. To that end we use the expression similar to eq.~(\ref{eq:ratios-meta2c-over-meta1c}) and fit our data to
\begin{eqnarray}
V_{12}^{\rm latt}(0)=V_{12}(0)^{{\rm cont}}  \left[ 1 + X_V \left({a\over a_{(\beta=3.9)}}\right)^2 \right] \,. \label{extrap-V}
\end{eqnarray}
That extrapolation is shown in fig.~\ref{fig:extrapo-V} and in the continuum limit we obtain
\begin{eqnarray}
V_{12}(0)= 0.32(6)\,,\qquad X_V=0.5(3)\,.
\label{final-V}
\end{eqnarray}
We see that the final error on $V_{12}(0)$ is rather large and the small effects due to fixing the charm quark mass and of the overall lattice spacing are completely immaterial at this stage. To account for a more important source of systematic uncertainty, we performed the continuum extrapolation by removing either the finest or the coarsest lattice and obtained $V_{12}(0)=0.31(8)$, and $V_{12}(0)= 0.34(8)$, respectively.  We can then take the spread of central values as an estimate of the error due to extrapolation to the continuum limit. To evaluate the impact of of higher excited states to our matrix element extraction, we also shortened the fitting region of $R_3(t)$, but keeping the points on the left of the plateaus (see fig.~\ref{fig:ratio_plateaux}) which are more likely to be sensitive to the higher excited states, and after the continuum extrapolation we obtain $V_{12}(0)=0.30(7)$. As our final result we quote 
\bea
V_{12}(0)= 0.32(6)(2)\,.
\eea
\subsection{Phenomenological discussion}

Let us first remind the reader of the value of the form factor $V_{21}(0)$, that parameterizes the hadronic matrix element describing $\psi (2S) \to \eta_c\gamma$ in a way completely analogous to eq.~(\ref{eq:def-vector-form-factor}). The decay branching fraction is given by,
\bea
B(\psi (2S) \to \eta_c\gamma) = \Gamma\bigl(\psi(2S)\bigr) \frac{8\alpha }{27 m_{\psi \left(2S\right)}^3 }\ \left(m_{\psi \left(2S\right)}^2-m_{\eta_{c}}^2\right) \left(m_{\psi \left(2S\right)}- m_{\eta_{c}}\right)^2 |V_{21}(0)|^{2},
\eea 
which can be combined with the experimental values~\cite{PDG},
\bea
&&m_{\eta_c}=2.9836(7)\ \gev\,,\quad m_{\psi(2S)} = 3.68611(1)\ \gev\,,\nn\\
&&\Gamma\bigl(\psi(2S)\bigr)=299(8)\ \kev\,,\quad B(\psi (2S) \to \eta_c\gamma) = 3.4(5)\times 10^{-3},
\eea
and $\alpha(m_c) = 1/134.$~\cite{alphaEM}, to get
\bea
V_{21}(0)=0.10(1)\,,
\eea
much smaller than $V_{12}(0)=0.32(6)$, that we obtained after extrapolating our lattice QCD results to the continuum limit.  
We attempted computing the form factor $V_{21}(0)$ at single lattice spacing and found it to be very small, consistent with zero within our error bars. 
It would take a huge statistics of the lattice data sample to be able to compute  $V_{21}(0)$ comparable with experimental accuracy. The value of $V_{12}(0)$, instead, was found to be large at every lattice spacing. 
Knowing that $\Gamma\bigl(\eta_c(2S)\bigr)=(11.4\pm 3.1)$~MeV, much larger than $\Gamma\bigl(\psi(2S)\bigr)$, the branching fractions of two decay modes become similar in size. More specifically, with  
\begin{align}
B(\eta_c (2S) \to J/\psi\gamma) = \Gamma\bigl(\eta_c(2S)\bigr) \frac{8\alpha }{9 m_{\eta_c \left(2S\right)}^3 }\ \left(m_{\eta_c \left(2S\right)}^2-m_{J/\psi}^2\right) \left(m_{\eta_c \left(2S\right)}- m_{J/\psi}\right)^2 |V_{12}(0)|^{2},
\end{align}
we get
\bea
\Gamma(\eta_c (2S) \to J/\psi\gamma)  = (15.7\pm 5.7)\ \kev\,,\qquad B(\eta_c (2S) \to J/\psi\gamma)  = 1.4(6) \times 10^{-3},
\eea
very similar to the measured $B(\psi (2S) \to \eta_c\gamma)$, and should be within reach at BESIII, KEDR, LHCb or Belle-2.

As we mentioned in introduction, the fact that $V_{12}(0)$ is much larger than $V_{21}(0)$ might come as a surprise because they differ by $1/m_c^2$-corrections and higher, that are expected to be reasonably small. However, from the quark model picture we know that a dominant contribution to these (hindered) transitions are absent due to orthogonality of the wave functions of initial and final states, and therefore the decay rates are almost entirely determined by the size of power corrections. In the effective field theoretical treatment of this problem, within pNRQCD, the terms ${\cal O}(1/m_c^2)$ have been identified~\cite{Brambilla:2005zw,Pineda:2013lta}. Unlike the allowed M1 transitions, such as $J/\psi \to \eta_c\gamma$ and $\psi(2S) \to \eta_c(2S)\gamma$, the hindered processes depend on the spin-spin interaction of the heavy quark potential which is precisely the one that affects differently $\eta_c(2S)\to J/\psi \gamma$ and $\psi (2S)\to \eta_c \gamma$,  cf. eq.~(\ref{pineda}). Assuming that all three operators ($ {\cal O}$), electromagnetic radius ($r^2$), typical velocity ($p^2/m_c^2$) and the spin operator ($V_{S^2}(\vec r)$) in eq.~(\ref{pineda}), satisfy ${}_{J/\psi}\langle {\cal O}\rangle_{\eta_c (2S) }= {}_{\eta_c}\langle {\cal O}\rangle_{\psi (2S)}$, from the difference of the two amplitudes we can estimate  ${}_{J/\psi}\langle V_{S^2}(\vec r) \rangle_{\eta_c(2S)}\equiv \langle V_{S^2}(\vec r) \rangle$. More specifically, 
\begin{align}
2 m_c \left( {V_{21}(0)\over m_{\psi(2S) }+ m_{\eta_c} } \right.&\left.  -  {V_{12}(0)\over m_{\eta_c(2S) }+ m_{J/\psi} } \right)
= \frac{2}{m_c^2} \langle V_{S^2}(\vec r) \rangle \left( {2 m_{\psi(2S)}\over m_{\psi(2S)}^2 - m_{\eta_c}^2} +  {2 m_{\eta_c(2S)}\over m_{\eta_c(2S)}^2 - m_{J/\psi}^2}\right)\nn\\
\Longrightarrow  &\quad   \langle V_{S^2}(\vec r) \rangle = m_c^3 \times (9.1 \pm 2.5)\times 10^{-3}\,,
\end{align}
which can be useful for phenomenology based on pNRQCD as a first estimate of the corresponding decay rates in the case of bottomia.

\section{Summary\label{sec:4}}

In this paper we made the first lattice computation of the form factor needed for a theoretical estimate of the radiative decay $\eta_c(2S)\to J/\psi \gamma$, and showed that its value is larger than the one entering the similar  $\psi(2S)\to \eta_c  \gamma$ decay. The explanation of that phenomenon can be understood in the pNRQCD description of these processes because they both involve a spin-spin interaction term which in the former decay enhances the decay rate and in the latter cancels the dominant power correction term. 

To be able to extract the matrix element for $\eta_c(2S)\to J/\psi \gamma$ on the lattice we needed to solve the GEVP and construct an interpolating field operator that couples mostly to $\eta_c(2S)$. We checked that in the continuum limit our result for $m_{\eta_c(2S)}/m_{\eta_c} = 1.23(2)$, is fully consistent with the measured $(m_{\eta_c(2S)}/m_{\eta_c})^{\rm exp.} \!= 1.22$. 
Our computations are made by using the (maximally) twisted mass QCD on the lattice by including $\nf =2$ dynamical light quarks and at four different lattice spacings. After taking the continuum limit and assuming the dependence on the light (sea) quark mass to be negligible, which we showed in our previous work to be the case for similar charmonium decays~\cite{Becirevic:2012dc}, we obtain that the decay width is
\bea
\Gamma(\eta_c(2S)\to J/\psi \gamma ) = (15.7\pm 5.7)\ \kev\,.
\eea
Since the width of $\eta_c(2S)$ is larger than that of $\psi (2S)$, the branching fraction we predict $B(\eta_c (2S) \to J/\psi\gamma)  = 1.4(6)\times 10^{-3}$ is very close to the experimentally established    
$B(\psi (2S) \to \eta_c\gamma)  =  3.4(5)\times 10^{-3}$, and could be within reach of BESIII, KEDR, LHCb, and/or Belle-2. 

The available experimental information on $B(\psi (2S) \to \eta_c\gamma)$ and $B(\psi (2S) \to \eta_c(2S)\gamma)$ together with our lattice results for the form factors describing $J/\psi\to \eta_c\gamma$~\cite{Becirevic:2012dc} and $\eta_c(2S)\to J/\psi\gamma$, confirm the expected pattern: the form factor is indeed large in the case of the allowed M1 transitions, while it is small for the hindered ones. Lattice QCD helped to solve the nonperturbative QCD effects where the experimental information was poor or not available, and we now have:
\bea
V_{11}(0)\equiv V(0)^{J/\psi\to \eta_c\gamma}_{\rm latt} = 1.92(3)(2)\,,\qquad V_{11}(0)\equiv V(0)^{\psi (2S)\to \eta_c (2S)\gamma} = 2.32(98)\,,\nn\\
V_{12}(0)\equiv V(0)^{\eta_c(2S)\to J/\psi\gamma}_{\rm latt} = 0.32(6)(2)\,,\qquad V_{21}(0)\equiv V(0)^{\psi (2S)\to \eta_c \gamma} = 0.10(1)\,. 
\eea

The value for the form factor $V_{12}(0)$ presented here can be improved by increasing statistics and by verifying that the form factor does not depend on the mass of the light sea quark. Furthermore a computation of  $V_{12}(0)$ by using a different lattice QCD discretization scheme would be very welcome as well. 

\section*{Acknowledgments}

We thank the members of the ETM Collaboration for making their gauge field configurations publicly available, A.~Pineda and A.~Vairo for discussions, and GENCI (2013-056808) for according us computing time at IDRIS Orsay where we performed numerical computations.

\vspace*{2cm}

\section*{Appendix}
The numerical code used for our physics projects is available under the GNU General Public License at {\tt https://code.google.com/p/nissa/}.

The code includes a possibility to consider different variations of the Wilson and Staggered Dirac operator. It is massively parallelized and supports both shared and/or distributed memory parallelism. 
It relies on MPI for communication among different computing nodes, and on a custom threading system for internal node parallelization. The former is especially suitable for Blue Gene/Q machine. Further optimization, for this particular architecture, and implemented in the code include: direct usage of the hardware communication layer (SPI) in key routines  to allow a better communication/computation overlap; specific data layout to optimize the memory access; usage of SIMD vectorized instructions (QPX) through interpretation of the registers' vector components as two complex
values from two virtual nodes. Currently, the maximum code performance, which is reached in the Dirac operator kernel, is around $25\div 40$\% of the machine's peak performance.


\begin{thebibliography}{99}

\bibitem{Gaiser:1985ix}
  J.~Gaiser, E.~D.~Bloom, F.~Bulos, G.~Godfrey, C.~M.~Kiesling, W.~S.~Lockman, M.~Oreglia and D.~L.~Scharre {\it et al.},
  Phys.\ Rev.\ D {\bf 34} (1986) 711.


\bibitem{Sucher:1978wq}
  J.~Sucher,
  Rept.\ Prog.\ Phys.\  {\bf 41} (1978) 1781.


\bibitem{Grotch:1984gf}
  H.~Grotch, D.~A.~Owen and K.~J.~Sebastian,
  Phys.\ Rev.\ D {\bf 30} (1984) 1924.

\bibitem{Godfrey:1985xj}
  S.~Godfrey and N.~Isgur,
  Phys.\ Rev.\ D {\bf 32} (1985) 189.


\bibitem{Shifman:1979nx}
  M.~A.~Shifman,
  Z.\ Phys.\ C {\bf 4} (1980) 345
   [Erratum-ibid.\ C {\bf 6} (1980) 282].

\bibitem{Khodjamirian:1983gd}
  A.~Y.~.Khodjamirian,
  Sov.\ J.\ Nucl.\ Phys.\  {\bf 39} (1984) 614
   [Yad.\ Fiz.\  {\bf 39} (1984) 970].
  
\bibitem{Beilin:1984pf}
V.~A.~Beilin and A.~V.~Radyushkin,
  Nucl.\ Phys.\ B {\bf 260} (1985) 61.

\bibitem{Mitchell:2008aa}
  R.~E.~Mitchell {\it et al.}  [CLEO Collaboration],
  Phys.\ Rev.\ Lett.\  {\bf 102} (2009) 011801
   [Erratum-ibid.\  {\bf 106} (2011) 159903]
  [arXiv:0805.0252 [hep-ex]].


\bibitem{Anashin:2014wva}
  V.~V.~Anashin {\it et al.}   [KEDR Collaboration],
  Phys.\ Lett.\ B {\bf 738} (2014) 391
  [arXiv:1406.7644 [hep-ex]].

\bibitem{Becirevic:2012dc}
  D.~Becirevic and F.~Sanfilippo,
  JHEP {\bf 1301} (2013) 028
  [arXiv:1206.1445 [hep-lat]].

\bibitem{Donald:2012ga}
  G.~C.~Donald {\it et al.}   [HPQCD Collaboration],
  Phys.\ Rev.\ D {\bf 86} (2012) 094501
  [arXiv:1208.2855 [hep-lat]].


\bibitem{Brambilla:2005zw}
  N.~Brambilla, Y.~Jia and A.~Vairo,
  Phys.\ Rev.\ D {\bf 73} (2006) 054005
  [hep-ph/0512369].


\bibitem{Pineda:2013lta}
  A.~Pineda and J.~Segovia,
  Phys.\ Rev.\ D {\bf 87} (2013) 7,  074024
  [arXiv:1302.3528 [hep-ph]].


\bibitem{A0}
  S.~Andreas, O.~Lebedev, S.~Ramos-Sanchez and A.~Ringwald,
  JHEP {\bf 1008} (2010) 003
  [arXiv:1005.3978 [hep-ph]];
R.~Dermisek and J.~F.~Gunion,
  Phys.\ Rev.\ D {\bf 81} (2010) 075003
  [arXiv:1002.1971 [hep-ph]];
F.~Domingo, U.~Ellwanger and M.~-A.~Sanchis-Lozano,
  Phys.\ Rev.\ Lett.\  {\bf 103}, 111802 (2009)
  [arXiv:0907.0348 [hep-ph]];
  R.~Dermisek, J.~F.~Gunion and B.~McElrath,
  Phys.\ Rev.\ D {\bf 76} (2007) 051105
  [hep-ph/0612031].


\bibitem{twbc}
  P.~F.~Bedaque,
  Phys.\ Lett.\ B {\bf 593} (2004) 82
  [nucl-th/0402051]; 
  G.~M.~de Divitiis, R.~Petronzio and N.~Tantalo,
  Phys.\ Lett.\ B {\bf 595} (2004) 408
  [hep-lat/0405002];
C.~T.~Sachrajda and G.~Villadoro,
  Phys.\ Lett.\ B {\bf 609} (2005) 73
  [hep-lat/0411033].


\bibitem{meissner}
F.~K.~Guo and U.~G.~Meissner,
  Phys.\ Rev.\ Lett.\  {\bf 109} (2012) 062001
  [arXiv:1203.1116 [hep-ph]];
  T.~Mehen and D.~L.~Yang,
  Phys.\ Rev.\ D {\bf 85} (2012) 014002
  [arXiv:1111.3884 [hep-ph]].


\bibitem{fr}
  R.~Frezzotti and G.~C.~Rossi,
  JHEP {\bf 0408}, 007 (2004)
  [arXiv:hep-lat/0306014].


\bibitem{boucaud}
P.~.Boucaud {\it et al.}  [ETM Collaboration],
  Phys.\ Lett.\ B {\bf 650} (2007) 304
  [hep-lat/0701012]; 
  Comput.\ Phys.\ Commun.\  {\bf 179} (2008) 695
  [arXiv:0803.0224 [hep-lat]].


\bibitem{ZZZ}
  M.~Constantinou {\it et al.}  [ETM Collaboration],
  JHEP {\bf 1008} (2010) 068
  [arXiv:1004.1115 [hep-lat]];
  arXiv:1201.5025 [hep-lat].

\bibitem{Blossier:2010cr}
  B.~Blossier {\it et al.}  [ETM Collaboration],
  Phys.\ Rev.\  D {\bf 82} (2010) 114513
  [arXiv:1010.3659 [hep-lat]].


\bibitem{GEVP}
  C.~Michael,
  Nucl.\ Phys.\ B {\bf 259} (1985) 58;
M.~Luscher and U.~Wolff,
  Nucl.\ Phys.\ B {\bf 339} (1990) 222;
B.~Blossier, M.~Della Morte, G.~von Hippel, T.~Mendes and R.~Sommer,
  JHEP {\bf 0904} (2009) 094
  [arXiv:0902.1265 [hep-lat]].


\bibitem{Gusken:1989ad}
  S.~Gusken, U.~Low, K.~H.~Mutter, R.~Sommer, A.~Patel and K.~Schilling,
  Phys.\ Lett.\ B {\bf 227} (1989) 266.



\bibitem{albanese}
  M.~Albanese {\it et al.}  [APE Collaboration],
  Phys.\ Lett.\ B {\bf 192} (1987) 163.


\bibitem{Liu:2012ze}
  L.~Liu {\it et al.}  [Hadron Spectrum Collaboration],
  JHEP {\bf 1207} (2012) 126
  [arXiv:1204.5425 [hep-ph]];
J.~J.~Dudek, R.~G.~Edwards, N.~Mathur and D.~G.~Richards,
  Phys.\ Rev.\ D {\bf 77} (2008) 034501
  [arXiv:0707.4162 [hep-lat]].


\bibitem{Becirevic:2013bsa}
  D.~Becirevic, G.~Duplancic, B.~Klajn, B.~Melic and F.~Sanfilippo,
  Nucl.\ Phys.\ B {\bf 883} (2014) 306
  [arXiv:1312.2858 [hep-ph]].


\bibitem{PDG} 
  K.~A.~Olive {\it et al.}  [Particle Data Group Collaboration],
  Chin.\ Phys.\ C {\bf 38} (2014) 090001.


\bibitem{DeTar:2012xk}
  C.~DeTar, A.~S.~Kronfeld, S.~H.~Lee, L.~Levkova, D.~Mohler and J.~N.~Simone,
  PoS LATTICE {\bf 2012} (2012) 257
  [arXiv:1211.2253 [hep-lat]].


\bibitem{lattice-radiative-1}
  J.~J.~Dudek, R.~G.~Edwards and D.~G.~Richards,
  Phys.\ Rev.\ D {\bf 73} (2006) 074507
  [hep-ph/0601137].


\bibitem{lattice-radiative-2}
  Y.~Chen, D.~-C.~Du, B.~-Z.~Guo, N.~Li, C.~Liu, H.~Liu, Y.~-B.~Liu and J.~-P.~Ma {\it et al.},
  Phys.\ Rev.\ D {\bf 84} (2011) 034503
  [arXiv:1104.2655 [hep-lat]].


\bibitem{alphaEM}
K.~Hagiwara, A.~D.~Martin, D.~Nomura and T.~Teubner,
  Phys.\ Rev.\ D {\bf 69} (2004) 093003
  [hep-ph/0312250];
  A.~A.~Pivovarov,
  Phys.\ Atom.\ Nucl.\  {\bf 65} (2002) 1319
   [Yad.\ Fiz.\  {\bf 65} (2002) 1352]
  [hep-ph/0011135];
J.~Erler,
  Phys.\ Rev.\ D {\bf 59} (1999) 054008
  [hep-ph/9803453].


\end{thebibliography}
\end{document}